\documentclass[a4paper,10pt,twoside]{cpc-hepnp}
\usepackage{CJK,upgreek,fancyhdr}
\usepackage{multicol}
\usepackage{graphicx}
\usepackage{booktabs}
\usepackage{sectsty}
\usepackage{amssymb,bm,mathrsfs,bbm,amscd}
\usepackage[tbtags]{amsmath}
\usepackage{lastpage}
\usepackage{color}
\usepackage[dvipdfmx, bookmarks=true, colorlinks,%
            citecolor=blue, linkcolor=blue, anchorcolor=blue,
            filecolor=blue,urlcolor=blue,hypertex, %
            breaklinks=true]{hyperref}

\begin{document}
\begin{CJK*}{GBK}{song}



\title{Improved predictions of phenomenological nuclear charge radius formulae with Bayesian optimization approach      \thanks{Supported by National Natural Science
               Foundation of China (12275115, 11447019)}}

\author{Song-Bo Zhao, \quad
        Lu Sun, \quad
        Cai-Xin Yua , \quad
        Ying-Chen Mao$^{1)}$\email{myc@lnnu.edu.cn}}%

\maketitle

\address{Department of Physics, Liaoning Normal University, Dalian 116029, China}

\end{CJK*}

\begin{abstract}
The model inputs play a key role in the performance of the Bayesian optimization approach. In this paper, we investigate the influence of the inputs on the improved predictions of phenomenological nuclear charge radius formulas using an approach combining those original formulas and the Bayesian neural network (BNN). We find that there is no improvement in predictions after the abnormal odd-even staggering effect of $^{181,183,185}$Hg is injected into the BNN, while the original phenomenological formulas themselves possess rich physical information or rigid constraints. It indicates the abundance and intensity of physical inputs affect the performance of the Bayesian optimization approach as well as the robustness of the BNN. We further demonstrate that, by ensuring that the number of neurons in the hidden layer is larger than the number of NN inputs, adding hidden layers into the BNN can significantly improve the predictions of nuclear charge radii formulas within the Bayesian optimization approach.
\end{abstract}




\maketitle

\begin{multicols}{2}

\section{Introduction\label{Sec:intro}}

Nuclear charge radius, one of most 
essential properties of a nucleus, is original reflection of nuclear charge density distribution, then plays a key role for ingoing experimental researches, such as electron scattering, $\alpha$ decay, and nuclear reactions~\cite{Avgoulea2011_JPG, Krieger2012_PRL, Ni2013_PRC}. Of course, it obviously affect our understanding of various nuclear structure characters~\cite{Norter.2009_PRL, Yang2016_PRL, Marsh2018_NP, DeGroote2019_NP, Kreim2014_PLB, Gorges2019_PRL},  as well as complicated dynamics phenomena ~\cite{Cheng2019_PRC, Xu2023_PLB, Yamamoto2017_PRC}. Besides a large number of experimental methods to directly measure the nuclear charge radius $R_{\rm{ch}}$, many chtheoretical methods have been developed to predict it. In general, there are mainly four types to describe the nuclear charge radius. The first one is the microscopic nuclear structure models, for instance, ab initio calculations with chiral effective field theory interactions, the various relativistic mean field (RMF) models, and the Skyrme-Hartree-Fock-Bogoliubov models, these models can well predict the nuclear properties of atomic nuclei both from the $\beta$-stability line to the driplines~\cite{Xia2018_ADNDT, Liu2024_ADNDT, An2020_PRC, An2022_CPC, Stoitsov2003_PRC, Liu2017_PRC, Liang2018_PRC}. The second one is semiempirical formulas, ranging from the simplest relations, $R \propto A^{1/3}$~\cite{Bohr1969_Book} and $R \propto Z^{1/3}$~\cite{Zeng1957_APS}, liquid drop models~\cite{Weizsacker1935_ZP, Brown1984_JPG}, to phenomenological formulae~\cite{NP1994_ZPA, Zhang2002_EPJA, Wang2013_PRC, Sheng2015_EPJA}. The third one is special methods, such as the methods like the Garvey-Kelson (GK) relations or the $\delta R_{in-jp}$ relations~\cite{Sun2014_PRC, Bao2016_PRC}, $\delta R_{nn}$ and $\delta R_{pp}$ relations~\cite{Bao2020_PRC, Li2023_CPC, Ma2021_PRC}, and the other special method directly derived from the quantum mechanical treatment of $\alpha$ decay with some approximations~\cite{Ni2013_PRC, Qian2013_PRC, Qian2014_PRC}. The last one, with rapidly developing in recent years, is machine learning methods~\cite{Carleo2019_RMP}, ranging from artificial neural networks (ANNs)~\cite{Akkoyun2013_JPG, Wu2020_PRC}, Bayesian neural networks (BNNs)~\cite{Utama2016_JPG, Dong2022_PRC, Dong2023_PLB}, Kernel ridge regression~\cite{Ma2022_CPC}, to Bayesian probability classifier~\cite{Ma2020_PRC, Tao2022_SSPMA}. 

The BNN as a \text{``}universal approximators\text{''} and powerful stochastic tool, have been successfully applied to investigate many nuclear problems, such as nuclear mass~\cite{Utama2016_PRC, Utama2017_PRC, Niu2018_PLB} and charge radii~\cite{Utama2016_JPG, Dong2022_PRC, Dong2023_PLB}, mass and charge distributions of fission yield~\cite{Wang2019_PRL, Qiao2021_PRC}, $\alpha$-decay and $\beta$-decay half-lives~\cite{Niu2019_PRC, Jin2023_PRC, Dong2023_PLB}, proton radius~\cite{Graczyk2014_PRC}, nuclear liquid-gas phase transition~\cite{Wang2020_PRR} and analysis of the Skyrme energy density functional ~\cite{Hizawa2023_PRC}. Both in Ref.~\cite{Utama2016_JPG} and Ref.~\cite{Dong2022_PRC, Dong2023_PLB}, a feed-forward BNN are used to simulate the original residuals of $R_{\rm{ch}} = R_{\rm{exp.}} -R_{\rm{th.}}$, but the later has different input features, besides the proton number $Z$ and the mass number $A$, four new inputs, i.e. the paring parameter $\delta$, the Casten factor $P$, the isospin factor $I^2$, and a $LI$ parameter which treats the ``abnormal" staggering behavior in $^{181,183,185}$Hg, are chosen as the final inputs. On the basis of a $50\%$ improvement compared to RMF for the former BNN, the latter, equipped with six inputs, achieves significantly enhanced predictions of $R_{\rm{ch}}$. It can effectively describe the odd-even staggering phenomena observed in the charge radii of the calcium~\cite{An2020_PRC} and potassium~\cite{An2022_CPC} isotopic chains.

Inspired by Ref.~\cite{Dong2023_PLB}, in order to investigate the optimization and extrapolating capabilities of the BNN methods as well as the robustness of BNN optimization methods to abnormal data ($LI$), we carry out our research works. In our work, the basic inputs of the BNN are initially the inherent parameters of the original phenomenological formulas, after which additional variables or parameters, such as $\delta$, $P$ and $LI$ are selected to be the BNN' inputs. We aim to improve our understanding of the Bayesian optimization approach and the physical mechanisms behind the original phenomenological formulas. Additional important research efforts include deciphering the underlying physical mechanisms contained in these formulas and examining the impact of NN hyperparameters, such as the quantity of hidden layers and neurons in each hidden layer, on the optimization performance of BNN.

This paper is organized as follows. A brief introduction of phenomenological formulas of nuclear charge radius and the Bayesian optimization approach are presented in Sec.~2. Results and discussions are presented in Sec.~3, followed by a summary in Sec.~4.

\section{Theoretical framework\label{Sec:theor}}

\subsection{Phenomenological formulae for nuclear charge radius}

Considering the physical character of nuclear saturation, and based on the successful liquid drop model, the charge radius of a nucleus is frequently defined as 
\begin{eqnarray}
R_{\rm{ch}}=r_A A^{1/3} \ , \label{eq:A1}
\end{eqnarray}
namely the $ A^{1/3}$ law, where $A$ is the mass number, and the relationship between $R_{\rm{ch}}$ and the root-mean-square (rms) charge radius $\langle r^2 \rangle^{1/2}$ is
\begin{eqnarray}
R_{\rm{ch}}=\sqrt{5/3}\langle r^2 \rangle^{1/2} \ . \label{eq:A2}
\end{eqnarray}
Here $r_A$ is set as different values for light nuclei and for very heavy nuclei, respectively, though fitting the experimental data~\cite{Angeli2013_ADNDT, Li2021_ADNDT}.

Zeng considered that the charge radius $R_{\rm{ch}}$ should be directly related to the charge number $Z$, then similarly to the $A^{1/3}$ law, he drew out the $Z^{1/3}$ law of charge radius of a nucleus~\cite{Zeng1957_APS, Zeng1975_APS}
\begin{eqnarray}
R_Z=r_Z Z^{1/3} \ , \label{eq:A3}
\end{eqnarray}
The parameter $r_Z$ remains almost a constant for $A \geq 40$.

Thereafter, there are mainly two lines to expand the phenomenological formulas of $R_{\rm{ch}}$. Along the $A^{1/3}$ law, taking the following effects into account, such as neutron excess (isospin effects), the shell correction, the effective numbers of valence particles (or holes), and pairing effects, there are four phenomenological formulas of $R_{\rm{ch}}$ as follows:
\begin{eqnarray}
R^{{\rm{NP}}94}_{\rm{ch}}={r_0}\left(1-a\frac{N-Z}{A}+\frac{b}{A}\right){A^{1/3}}\ , \label{eq:A5}
\end{eqnarray}
\begin{eqnarray}
R^{{\rm{ELD}}13}_{\rm{ch}}={({r_0}+{r_1}{A^{-2/3}}+{r_2}{A^{-4/3}})}{A^{1/3}}\ , \label{eq:A6}
\end{eqnarray}
\begin{eqnarray}
R^{{\rm{Wang}}}_{\rm{ch}}={r_0}{A^{1/3}}+{r_1}{A^{-2/3}}+{r_s}I\left({1-I}\right)+{r_d}{{\Delta E}/A}\ , \label{eq:A7}
\end{eqnarray}
\begin{eqnarray}
R^{{\rm{Sheng}}}_{\rm{ch}}={r_0}\left({1-a\frac{N-Z}{A}+\frac{b}{A}+c\frac{P}{A}+d\frac{\delta }{A}}\right){A^{1/3}}\ , \label{eq:A8}
\end{eqnarray}
where $I=\frac{N-Z}{A}$ is isospin parameter, $\Delta E$ denotes the shell corrections of the nuclei from the WS$^{*}$ mass model~\cite{Wang2014_PLB}. The Casten factor $P=\frac{N_p N_n}{N_p + N_n}$, where $N_p$ and $N_n$ represent the numbers of valence protons and neutrons, respectively, the following magic numbers $Z_M=2,6,14,28,50,82,(114)$, and  $N_M=2,6,14,28,50,82,126,(184)$ are employed to make the best fit~\cite{Angeli1991_JPG, Dieperink2009_EPJA}. 

Along another line, the $Z^{1/3}$ law, Zhang and his collaborators develop Zeng's law \cite{Zhang2002_EPJA}, then a new formula can be written as
\begin{eqnarray}
R^{{\rm{Zhang}}}_{\rm{ch}}=r_Z \left(1+\frac{5\beta}{8\pi^2}\right)\left(1+b\frac{N-N^\ast}{Z}\right)Z^{1/3}\ , \label{eq:A9}
\end{eqnarray}
where $\beta$ is the quadrupole deformation, which is taken from~\cite{Moller2016_ADNDT} in the present work. All coefficients appear in Eqs.~(4)-(8) are listed in Table 1.

\end{multicols}
\begin{center}
\renewcommand\arraystretch{1.5}
\tabcaption{ \label{t:1}
Coefficients and the root-mean-square deviations $\sigma^{{\rm{CR}}13+21}_{{\rm{rms}}}$ for the six phenomenological nuclear charge radius formulas. Experimental data are taken from ~\cite{Angeli2013_ADNDT, Li2021_ADNDT}, with proton number $Z \geq 8$ and neutron number $N \geq 8$.}
\footnotesize
\begin{tabular*}{0.95\textwidth}{l@{\extracolsep{\fill}}lcc}
\hline
\hline
Formula &Parameters &  $\sigma_{\rm{rms}}^{\rm{CR13+21}}$~(fm) \\
\hline
$R^{{\rm{NP}}94}_{\rm{ch}}={r_0}\left({1-a\frac{{N-Z}}{A}+\frac{b}{A}}\right){A^{1/3}}$      & ${r_0}=1.240$~\rm{fm}, $a=0.191$, $b=1.646$~\cite{NP1994_ZPA}  & 0.0538  \\
$R^{{\rm{Zhang}}}_{\rm{ch}}=r_Z (1+5\beta/8\pi^2)\left[1+b(N-N^\ast)/Z\right]Z^{1/3}$ &
   ${r_{Zd}}=1.631$~\rm{fm}, $b=0.062$~\cite{Zhang2002_EPJA}   & 0.0489 \\
$R^{{\rm{ELD}}13}_{\rm{ch}}=\left({{r_0}+{r_1}{A^{-2/3}}+{r_2}{A^{-4/3}}}\right){A^{1/3}}$  &
${r_0}=0.9071$~\rm{fm}, ${r_1}=1.105$~\rm{fm}, ${r_2}=-0.548~\rm{fm}$~\cite{Angeli2013_ADNDT}&    0.0551  \\
$R^{{\rm{Wang}}}_{\rm{ch}}={r_0}{A^{1/3}}+{r_1}{A^{-2/3}}+{r_s}I\left({1-I}\right)+{r_d}{{\Delta E}/A}$& ${r_0}=1.2260$~\rm{fm}, ${r_1}=2.86$~\rm{fm},  &   0.0213 \\
$~~~$ & ${r_s}=-1.09~{\rm{fm}}$, ${r_d}=-1.09~\rm{MeV}^{-1}~\rm{fm}$~\cite{Wang2013_PRC}& $~~~$ \\
$R^{{\rm{Sheng}}}_{\rm{ch}}={r_0}\left({1-a\frac{{N-Z}}{A}+b\frac{1}{A}+c\frac{P}{A}+d\frac{\delta }{A}}\right){A^{1/3}}$ 
&${r_0}=1.2321~\rm{fm}$, $a=0.1534$, &  0.0314  \\
~~~&$b=1.3358$, $c=0.4317$, $d=0.1225$~\cite{Sheng2015_EPJA}  \\
\bottomrule
\end{tabular*}
\end{center}
\begin{multicols}{2}

\subsection{Bayesian optimization method}
The BNN~\cite{Neal1996_BOOK}, combining an artificial NN with the Bayesian statistical theory, is a universal optimization method capable of accurately fitting any analytical function. In our work, we employed a feed-forward neural network (FFNN) with a single or multi-hidden layer, connecting the input and output layers, as shown in Fig. 1. If there are three hidden layers, the architecture can be denoted as I-J-K-L-1 FFNN, where $I$ represents the number of inputs and $J$, $K$, and $L$ correspond respectively to the numbers of neurons in the three feed-forward hidden layers. We will let $M$ denote the number of layers in the network, thus $M = 5$ in the I-J-K-L-1 FFNN. labeling layer 1 as $L1$, which is just the input layer, and layer $L5$ as the output layer.

\begin{center}
\vspace{1mm}
\centering
\includegraphics[scale=0.08]{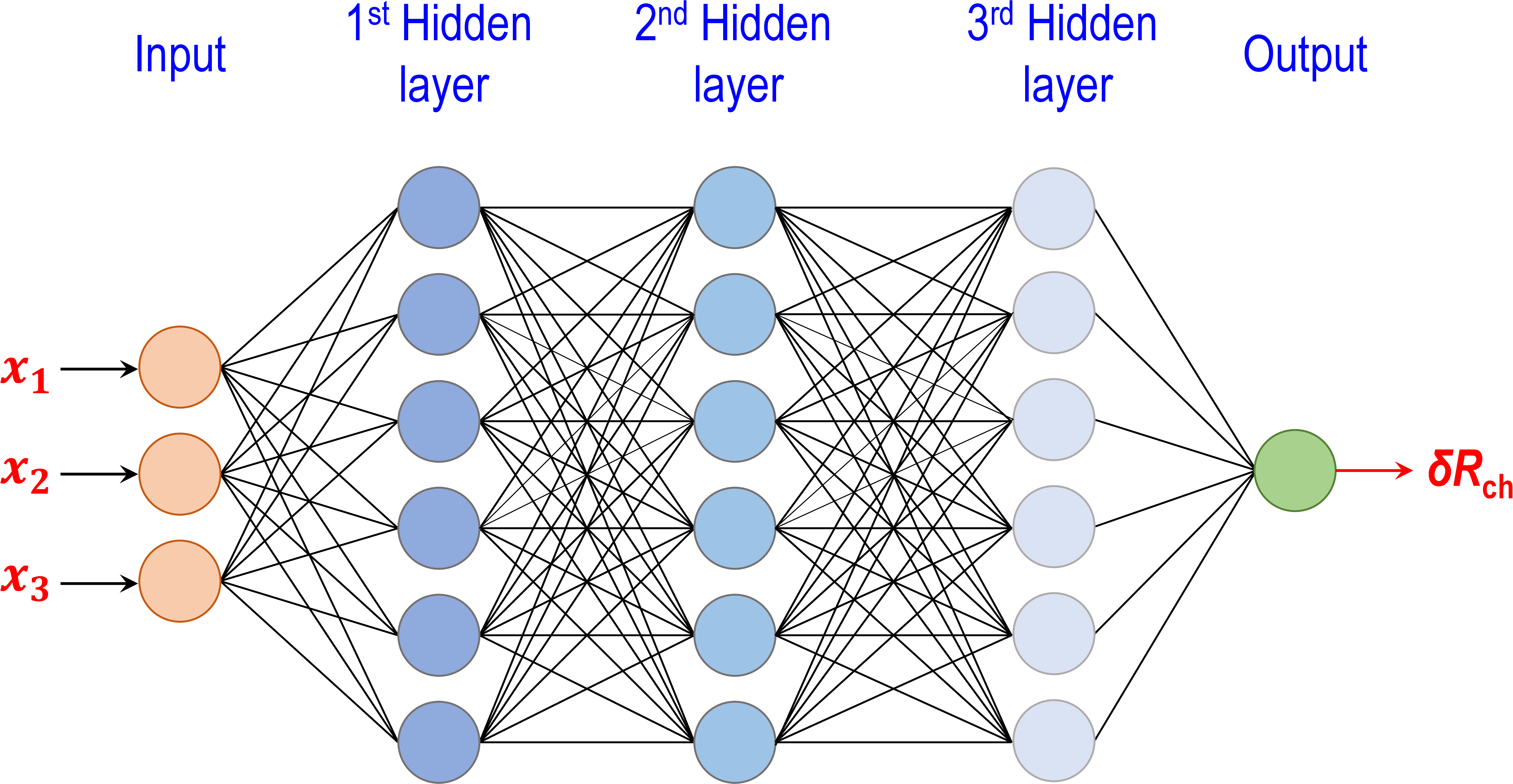}
\vspace{2mm}
\figcaption{\label{FFNN:1}
Illustration of the functional structure of a feed-forward neural network with multi-hidden layers for fitting the residuals of nuclear charge radii with different input parameters in the input layer. Note that the bias of each layer is not drawn in this diagram. The number of hidden layers is changed according to the needs of our study.}
\end{center}

The input and output of the $j$th neuron in layer $L2$, namely the first hidden layer, are given by 
\begin{equation}
x_{j}^{2}= b_{j}^{2} +  \sum_{i= 1}^{I} \left ( \omega _{ji}^{2} \times x_{i}^{1}  \right ),~~~~j=1,2\dots ,J 
\end{equation}
and
\begin{equation}
y_{j}^{2} =f^{2} \left ( x_{j}^{2}  \right ) ,~~~~j=1,2,\dots ,J,
\end{equation}
similarly, the input and output of the $k$th neuron 
and $l$th neuron in layer $L3$ and $L4$, respectively, are

\begin{equation} 
\begin{split}
 &x_{k}^{3}=b_{k}^{3}+\sum_{j=1}^{J}\left(w_{k j}^{3} \times 
y_{j}^{2}\right),
\quad k=1,2, \ldots, K\\
 &  y_{k}^{3}=f^{3}\left(x_{k}^{3}\right),
\end{split}
\end{equation}
and
\begin{equation}
\begin{split}
 &x_{l}^{4}=b_{l}^{4}+\sum_{k=1}^{K}\left(w_{l k}^{4} \times 
y_{k}^{3}\right),
\quad l=1,2, \ldots, L\\
 &  y_{l}^{4}=f^{4}\left(x_{l}^{4}\right),
\end{split}
\end{equation}
The output layer, i.e., layer $L5$ can be written as
\begin{equation}
 y_{m}^{5}=b_{m}^{5}+\sum_{l=1}^{L}\left(w_{m l}^{5} \times y_{l}^{4}\right), ~~~~m=1,2, \ldots, M,
\end{equation}
where, actually $m =1$ because that the FFNN has one
output in this work. $x_{i}^{1} \left ( i=1,2,\dots ,I \right )$ in Eq. (1) corresponds to the 
input variables. According to Eq. (4), the $k$th and
$l$th neurons in layer $L3$ and $L4$ are connected with weights $\omega _{lk}^{4}$,
and similar expressions for other weights.
The threshold of the $j$th neuron of $Lj$ layer 
is determined by its biases. The transfer functions $fs$ are chosen as ReLU functions. In this work, 
the output layer provides calibrated residuals after BNN' prediction, which optimizing the target data $\delta R_{\rm{ch}} =R_{\rm{exp}} -R_{\rm{th}}$,
i.e., the raw residuals between experimental
and theoretical nuclear charge radii. 

In the Bayesian optimization method, the posterior distribution
\begin{equation}
 p\left ( \omega |x,t \right ) =\frac{p\left ( x,t|\omega  \right ) p\left (  \omega \right ) }{p\left ( x,t \right ) } 
\end{equation}
can be calculated by Bayes's theorem~\cite{Neal1996_BOOK}, which is used to make predictions. Here $p\left ( \omega \right) $ is the prior distribution of 
the free parameters $\omega $ (such as bias, and weights) of the neural
network, $p\left ( x,t|\omega  \right ) $ is the likelihood 
function, $p\left ( x, t\right) $ is the marginal likelihood 
which can always be neglected because
it does not contain the information of parameters, 
and $t$ is the set of target data, namely the raw residuals $\delta R_{\rm{ch}} $. All the model parameters are assumed to be independent in this work. 

Following the standard practice, we assume the likelihood function obey Gaussian distribution,
i.e., $p\left (x,t|  \omega \right) =\rm{exp}\left ( -\chi ^{2}/2  \right ) $ , 
where the function $\chi ^{2} \left ( \omega  \right ) $ is given by
 \begin{equation}
\chi^{2}(\omega)=\sum_{i=1}^{N}\left(\frac{t_{i}-f\left(x_{i}^{1}, \omega\right)}{\Delta t_{i}}\right)^{2} ,\\
\end{equation}
where $N$ is the number of training data. Similarly, 
the prior distributions $p\left ( \omega  \right )$  
is assigned as a zero mean Gaussian function,
and modeled by a $\gamma $ distribution. After training within the BNN, the free parameters obey the posterior distribution, then the target data is given by
\begin{equation}
\left\langle f_{n}\right\rangle=\int f\left(x_{i}, 
\omega\right) p(\omega \mid x, t) d \omega ,\\  
\end{equation}
then the statistical uncertainty of BNN prediction can be written as
\begin{equation}
 \triangle f_{n}=\sqrt{\left\langle f_{n}^{2}\right\rangle-\left\langle f_{n}\right\rangle^{2}},
\end{equation}
where $\left\langle f_{n}^{2}\right\rangle$ can be obtained with the same procedure presented in Eq. (16).

\section{Results and discussion\label{Sec:results}}

Similarly to Ref.~\cite{Ma2022_CPC}, we study those nuclei with $Z \geq 8$, and $N \geq 8$ in this work. For the training set, we use the 781 experimental data listed in Ref.~\cite{Angeli2004_ADNDT}. The more recent experimental data ~\cite{Angeli2013_ADNDT, Li2021_ADNDT}, 232 data for $R_{\rm{ch}}$  with the same criterion, are selected as the validation set to test the predictive capabilities of BNN optimization method. The root-mean-square deviation (RMSD) $\sigma_{\rm{rms}} $ is used to quantify the predictive ability of different models after BNN optimization
\begin{equation}
 \sigma  _{\rm{rms}} =\sqrt{\frac{1}{N}\sum_{i=1}^{N} (R_{i}^{\rm{exp}} -R_{i}^{\rm{th}} )^{2}  }, 
\end{equation} 
where $N$ is the total number of charge radii collected in the entire set. 

\subsection{Global optimization of the Bayesian optimization method and the influence of abnormal data}

In order to compare with other works, we first carry out the calculations with a simple hidden layer. Referring to Ref.~\cite{Dong2023_PLB}, for the sake of easy description, we set D$i$ to denote those models combining original phenomenological formulas with $i$-inputs in the BNN models. Those inputs for five models are listed in the second column of Table 2.

Based on the raw residuals, namely the residuals between experimental and theoretical (original formulas) charge radii, the BNN method can refine the theoretical results and  obtain the calibrated $R_{\rm{ch}}$; hence, the rms deviations $\sigma_{\rm{post}}$ can be obtained. Let the rms deviations $\sigma_{\rm{pre}}$ correspond to the original formulas. In order to measure the improvement of the BNN method, the relative difference in RMSDs is written by $\Delta\sigma = \frac{\sigma_{\rm{pre}}-\sigma_{\rm{post}}}{\sigma_{\rm{pre}}}$, and the three quantities are also listed in Table 2.

\end{multicols}
\begin{center}
\tabcaption{ \label{t:2}
The rms deviations $\sigma_{\rm{pre}}$, $\sigma_{\rm{post}}$, and the relative difference in rms deviations $\Delta \sigma$ of nuclear charge radii $R_{\rm{ch}}$ for the training, validation, and entire sets, respectively. Here, the inputs only includes those input variables of the BNN other than ordinary variables $Z$ and $A$, otherwise where will be blank.}
\centering
\scalebox{0.92}{
\begin{tabular}{c l c c c c c c c c c c c}
\hline\hline
\multicolumn{1}{c}{}&~&\multicolumn{1}{c}{}&\multicolumn{3}{c}{Training set} &\multicolumn{3}{c}{Validation set} &~&\multicolumn{3}{c}{Entire set} \\
\cline{3-5}\cline{7-9}\cline{11-13}
\raisebox{1ex}[1ex] {Model} &\raisebox{1ex}[1ex] {Inputs} &$\sigma_{\rm{pre}}/\rm{fm}$ &$\sigma_{\rm{post}}/\rm{fm}$ &$\Delta \sigma /\%$  &~&$\sigma_{\rm{pre}}/\rm{fm}$ &$\sigma_{\rm{post}}/\rm{fm}$ &$\Delta \sigma /\%$ &~&$\sigma_{\rm{pre}}/\rm{fm}$ &$\sigma_{\rm{post}}/\rm{fm}$ &$\Delta \sigma /\%$  \\ \hline
NP94&$~$&0.0557 & 0.0302& 45.8 & &0.0468 & 0.0316 & 32.5 & &0.0538  & 0.0306 & 43.2  \\
~&$(I)$&~& 0.0270 & 51.5 & &~& 0.0305 & 34.8 & &~& 0.0278   & 48.2 \\
~&$(I, LI)$&~& 0.0315 & 43.4& &~& 0.0315 & 32.7 & &~& 0.0315  & 41.4 \\ \cline{2-13}
Zhang&$~$&0.0493& 0.0257 & 47.9& &0.0475& 0.0217 & 54.3 & &0.0489 & 0.0248  & 49.2  \\
~&$(\beta)$&~&0.0239& 51.6 & &~& 0.0203& 57.3 & &~& 0.0231 & 52.7\\
~&$(\beta, \eta, \eta^{\ast})$&~& 0.0218 & 55.8 & &~& 0.0179 & 62.3 & &~& 0.0210 & 57.2\\
~&$(\beta, \eta, \eta^{\ast}, LI)$&~& 0.0221  & 55.2 & &~& 0.0188& 60.4 & &~& 0.0214  & 56.3\\ \cline{2-13}
ELD13&$~$&0.0513 & 0.0331 & 35.5& & 0.0663  & 0.0345  & 48.0 & &0.0551   & 0.0334    & 39.4 \\
~&$(LI)$&~& 0.0296 & 42.3 & &~& 0.0313   & 52.8 & &~& 0.0300  & 45.6 \\ \cline{2-13}
Wang&$~$&0.0222& 0.0214 & 3.6& &0.0179  & 0.0175& 2.2 & & 0.0213   & 0.0206 & 3.5\\
~&$(I)$&~& 0.0203 & 8.6 & &~& 0.0159   & 11.2 & &~& 0.0194 & 9.2   \\
~&$(I, \Delta{E})$&~& 0.0193 & 13.1 & &~& 0.0150 & 16.2 & &~& 0.0184    & 13.7   \\
~&$(I, \Delta{E}, LI)$&~& 0.0191 & 14.0 & &~& 0.0150 & 16.2 & &~& 0.0182    & 14.4  \\ \cline{2-13}
Sheng&$~$&0.0320  & 0.0244 & 23.8 & & 0.0291  & 0.0209  & 28.2 & &0.0312   & 0.0233    & 25.1 \\
~&$(I)$&~& 0.0216 & 32.5 & &~& 0.0189  & 35.1 & &~& 0.0213    & 31.8  \\
~&$(I, P, \delta)$&~& 0.0204 & 36.2& &~& 0.0189  & 35.1 & &~& 0.0195    & 37.4 \\
~&$(I, P, \delta, LI)$&~& 0.0210 & 34.4 & &~& 0.0190& 34.7 & &~& 0.0201  & 35.7 \\
\hline \hline
\end{tabular}
}
\end{center}
\begin{multicols}{2}

As can be seen from columns 9-11 of Table 2, the predictions of all five phenomenological models have improved after the optimizations of the BNNs. Here, the optimization for the Zhang02 formula is the largest, although its $\sigma_{\rm{pre}}$ value is not the largest. Meanwhile, the optimization of Wang formula, which directly takes into account 
the isospin effect and shell correction, has the 
smallest enhancement, which is mainly due 
to the fact that the RMSD of its original formula is already very low.
On the other hand, when the abnormal data input of Hg isotopes is not taken into account, we can see that with the increase of the physical information contained in the original phenomenological formula, i.e., with the increase of the BNN inputs, the optimization capability of all models for the charge radius is improved more substantially.
After considering the abnormal data input of ${}^{181,183,185}$Hg, 
we find that the optimization ability of the other three models, 
except for the Wang-BNN model and the ELD13-BNN model, 
decreases to some extent. A more careful comparison reveals 
that the enhancement of the ELD13-BNN model stems more
from the simplicity of its model inputs, and the input 
of abnormal data on Hg isotopes rather reduces the 
optimization capability of the BNN as more physical information is
injected, which is not entirely consistent with the 
conclusions obtained in Ref.~\cite{Dong2022_PRC, Dong2023_PLB} based on the NP94 formula.

Fig. 2 shows the comparison between the Wang model
and Sheng model with and without considering the abnormal
data input (labeled as Wang-D5 and Sheng-D6, respectively). 
Comparing Fig. 2(a) and (b), and Fig. 2(c) and (d), respectively, we can see that the two models that considered the input of abnormal data for Hg isotopes did not show significant improvement compared to the two models that were not considered. Therefore, we conclude that a small portion of abnormal data input did not have a general improvement effect on the optimization capability of BNNs. The more complex the phenomenological formula is, i.e. the more physical information it contains, or the stronger the physical constraints, the less significant the optimization improvement of BNNs on the original formula.

\end{multicols}
\begin{center}
\vspace{0.5mm}
\centering
\includegraphics[width=0.7\textwidth]{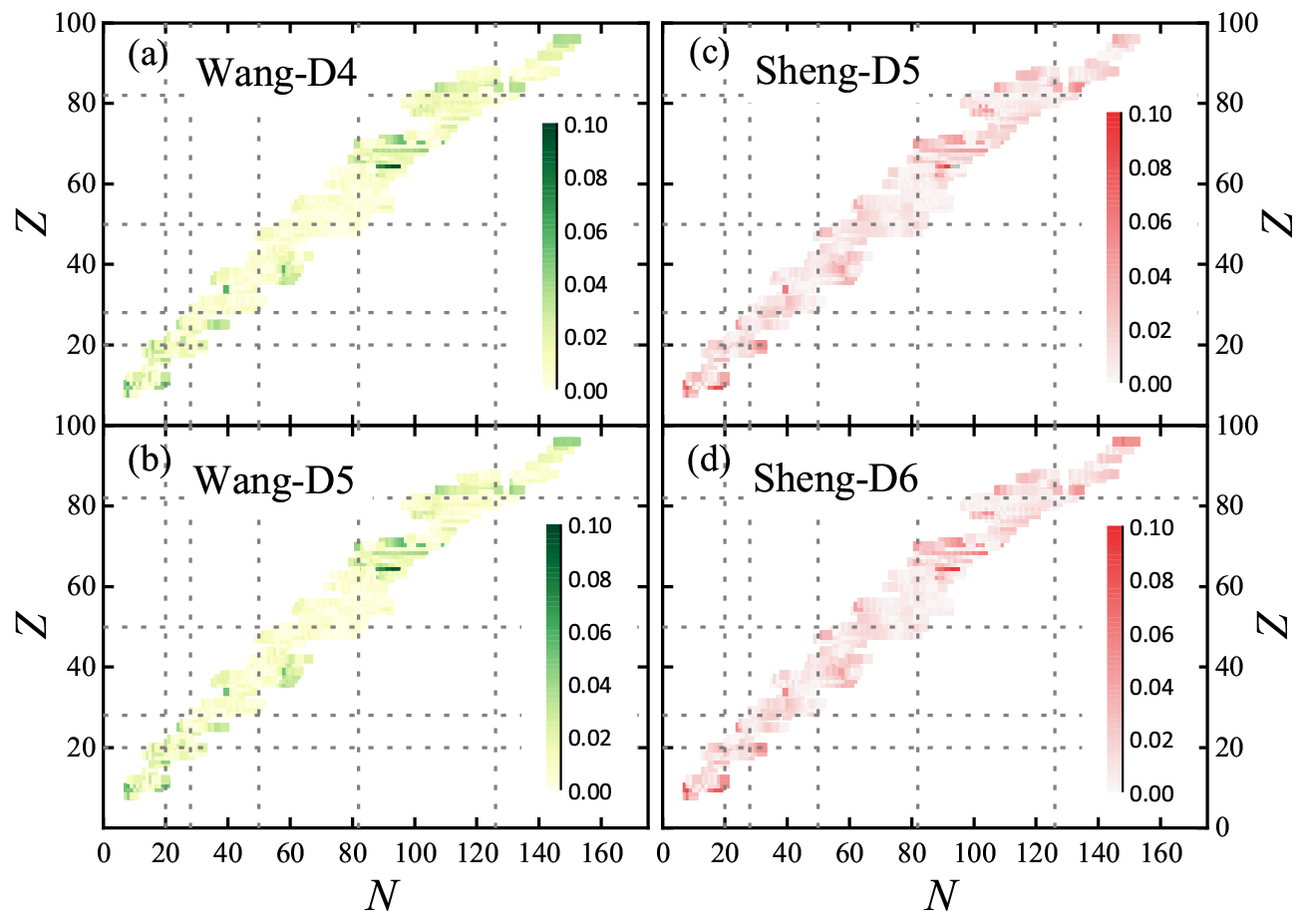}
\vspace{0.5mm}
\figcaption{\label{fig:2}
Predictions for the residuals (in units of fm) between the experimental $R_{\rm{ch}}$ and the Wang-D4 model (a), Wang-D5 model (b), Sheng-D5 model (c), and Sheng-D6 model (d), respectively. The inputs of the BNN models in the lower panel include the abnormal data on $^{181,183,185}$Hg.}
\end{center}
\begin{multicols}{2}

Similarly to the treatment of Hg isotope abnormal data in 
Ref.~\cite{Dong2023_PLB}, we screened all experimental nuclear charge radii~\cite{Angeli2013_ADNDT, Li2021_ADNDT} and performed a global search for abnormal data with odd-even shape staggering. We label a nuclide as abnormal data when the radius difference between it and its neighboring isotopes is greater than 0.01~fm. After counting the 1013 nuclides with experimental data, there are 16 nuclides with odd-even shape staggering, i.e., $^{18}$Ne, ${}^{21}$Na, ${}^{23,24}$Mg, ${}^{42-44}$Ca, ${}^{45,46}$Ti, ${}^{54}$Fe, ${}^{185}$Pt, and ${}^{181-185}$Hg, among which this shape staggering phenomenon is 
especially significant near ${}^{181-185}$Hg, where the nuclear charge radii between two neighboring isotopes, the difference is greater than or close to 0.04~fm. In order to investigate the influence of the abnormal data about the odd-even shape staggering, we selected the three models, namely, ELD13-BNN, Wang-BNN, and Sheng-BNN to carry out the following calculations, i.e., comparing the different strategies about the inputs of those abnormal data: (1) as in Ref.~\cite{Dong2023_PLB}, only the $LI$s of ${}^{181,183,185}$Hg are set to 1 (labeled as Hg$-3$); (2) the $LI$s of ${}^{181-185}$Hg is set to 1 (labeled as Hg$-5$); (3) the $LI$s of all the afore mentioned 16 nuclides is set to 1 (labeled as all), and the calculation results are displayed in Table 3. 

As can be seen from Table 3, except for the ELD13-BNN model, the inputs 
of the three different abnormal data have a very small 
effect on the Wang-BNN and Sheng-BNN models. For example, for the
validation set of the Wang-BNN model, the maximal difference 
of the rms deviation among the three cases is 
only 0.003~fm, while for the Sheng-BNN model, the rms deviation 
maximum difference is 0.01~fm, and it is better to set five Hg isotopes, i.e. ${}^{181-185}$Hg as abnormal data than to set only ${}^{181,183,185}$Hg as abnormal data.

\end{multicols}
\begin{center}
\tabcaption{ \label{t:3}
The RMSDs $\sigma_{\rm{pre}}$, $\sigma_{\rm{post}}$, and the relative difference in RMSDs $\Delta \sigma$ of nuclear charge radii $R_{\rm{ch}}$ predicted by ELD13-BNN, Wang-BNN, and Sheng-BNN, respectively. In the calculations, the inputs of three BNNs include three different treatment cases of abnormal data.}
\centering
\scalebox{0.95}{
\begin{tabular}{c c c c c c c c c c c c c}
\hline \hline
\multicolumn{1}{c}{}&\multicolumn{1}{c}{}&\multicolumn{3}{c}{Training set} & &\multicolumn{3}{c}{Validation set}  & &\multicolumn{3}{c}{Entire set}\\
\cline{3-5}\cline{7-9}\cline{11-13}
\raisebox{1ex}[1ex] {Model} &\raisebox{1ex}[1ex] {$LI$} &$\sigma_{\rm{pre}}/\rm{fm}$ &$\sigma_{\rm{post}}/\rm{fm}$ &$\Delta \sigma /\%$  & &$\sigma_{\rm{pre}}/\rm{fm}$ &$\sigma_{\rm{post}}/\rm{fm}$ &$\Delta \sigma /\%$  & &$\sigma_{\rm{pre}}/\rm{fm}$ &$\sigma_{\rm{post}}/\rm{fm}$ &$\Delta \sigma /\%$ \\
\hline
ELD13&$\rm{Hg}-3$&0.0513 & 0.0296 & $\centering{42.3}$& & 0.0663  & 0.0313  & 52.8& &0.0543   & 0.0294 & 45.8  \\
&$\rm{Hg}-5$& & 0.0320 & 37.6 & & & 0.0329   & 50.4 & & & 0.0315  & 41.9 \\ 
&$\rm{All}$& & 0.0311 & 39.4 & & & 0.0345   & 48.0 & & & 0.0312  & 42.5 \\ \cline{2-13}
Wang&$\rm{Hg}-3$&0.0222& 0.0191 & 14.0& &0.0179  & 0.0150& 16.2& & 0.0213   & 0.0182 & 14.4 \\
&$\rm{Hg}-5$& & 0.0194 & 12.6 & & & 0.0150 & 16.2 & & & 0.0185    & 13.4  \\
&$\rm{All}$& & 0.0188 & 15.3 & & & 0.0153 & 14.5 & & & 0.0181    & 15.2  \\ \cline{2-13}
Sheng&$\rm{Hg}-3$&0.0320  & 0.0207 & 35.3 & & 0.0291  & 0.0188  & 35.4  & &0.0299   & 0.0200 & 33.3  \\
&$\rm{Hg}-5$& & 0.0214 & 33.1& & & 0.0178  & 38.8 & & & 0.0202    & 32.5 \\
&$\rm{All}$& & 0.0213 &33.4& & & 0.0182& 37.5 &  & & 0.0205  & 31.7 \\
\hline \hline
\end{tabular}
}
\end{center}
\begin{multicols}{2}

Combined with Ref.~\cite{Dong2023_PLB},
we believe that for the NNs, their optimization efficiency for a certain phenomenological model depends on the selection of BNN input variables. Even though a certain original model contains relatively little physical information, as long as the inputs of the NNs containing more physical information are optimized using the BNN method, an improvement in the prediction accuracy will be achieved, such as the ELD13 formula. As more physical information is injected into the original phenomenological models, such as isospin asymmetry, nucleon ratio on the $\beta$-stability line, Casten factor (relating to the shell closure effects), and paring parameters, the impact of abnormal data input on the optimization capability of the BNNs becomes less significant, such as the Sheng formula. In addition, if the original empirical model contains strong physical constraints, such as Wang formula directly including shell correction energy, the impact of abnormal data input on the optimization capability of BNNs is likewise insignificant.

\subsection{Extrapolating capabilities of Bayesian optimization method}

As can be seen in columns 3-5, 6-8 of Table 2, there is
a relatively obvious overfitting of NP94-BNN and ELD13-BNN. Our understanding of this is that since there are only 1013 experimental data available for the nuclear charge radius, i.e., the dataset is too small, ordinary inputs to the BNN (e.g., only two inputs, Z and A), will mask some of the physical information. Compared with the ELD13-BNN model, the NP94-BNN model has better prediction accuracy on the training set than the validation set, which means that the latter has stronger extrapolation capability than the former. However, after considering abnormal data input, the prediction accuracy obtained on the two datasets is similar. A more careful comparison shows that considering abnormal data input does not change the optimization of the training set by the ELD13-BNN model compared to the validation set. Based on the above conclusions, it indicates that abnormal data input can affect to some extent the extrapolation capability of the NN model, which also reflects the importance of the model inputs in the NN training from another perspective.

For the other three models, we can find that the Zhang-BNN model also has some overfitting phenomena. Compared with them, the Wang-BNN and Sheng-BNN models have better extrapolation capabilities. It is worth pointing out that although the Wang-BNN model has better prediction accuracy, the extrapolation ability of the Sheng-BNN model is relatively better than that of the Wang-BNN model. This can be attributed to the fact that when we used the Wang-model for calculations, we used the shell correction energy provided by WS4~\cite{Wang2014_PLB}, while the original Wang formula used WS3.3 data.~\cite{Wang2010_PRC81, Wang2010_PRC82}. The shell correction is very important for determining the structural information 
such as the nuclear charge radius, and the shell correction
energies given by different methods differ greatly, so the
accurate calculation of the shell correction energy will 
seriously affect the raw residuals, therefore for the
validation set that contains the new data of $R_{\rm{ch}}$ ~\cite{Li2021_ADNDT}, the extrapolation of the Wang-BNN model is weaker than that of the Sheng-BNN model. Combined with Table 2, we can find that the RMSD calculated by original Wang formula is significantly smaller than that of original Sheng formula, which implies that the physical constraints on the model by directly including the shell correction energy in the phenomenological formula are stronger than those of Sheng formula that takes into account the Casten factor correction for shell closure effects. The extrapolation capability of the Wang-BNN model is more flabby than that of the Sheng-BNN model, which seems to mean that the stronger the physical constraints of the original formula, the less obvious the relative optimization of BNN on it. At the same time, its extrapolation capability is relatively lower, that is, the robustness of the model is relatively weaker.

For the Zhang, Wang, and Sheng formulas, since they all
contain the mass number $A$ and the isospin parameter $I$, we can be informed of the extrapolation capability of the BNN models corresponding to these original formulas from the variation of the RMSDs with the mass number $A$ and the isospin asymmetries $|N-Z|$. As can be seen from the left column of Fig. 3, the RMSDs obtained from the D$i_{\rm{max}}$ model exhibit the less fluctuation and the relative most stable mass dependence compared to the other models, especially the Zhang-D6 and Sheng-D6 models, from which the RMSDs obtained from the light and heavy nuclei are almost always the best. For the Wang-D5 model, although the RMSDs of heavy and medium heavy nuclei perform the best, their dependence on mass is not stable and behaves larger fluctuation compared to the other two models, which further reflects the relatively weak extrapolation capability of the model. As can be seen from the right column of Fig. 3, the RMSDs vary similarly with the isospin asymmetry $|N-Z|$, and the RMSDs obtained from the D$i_{\rm{max}}$ model as a whole almost display the most stable isospin asymmetry dependence.

\end{multicols}
\vspace{0.5mm}
\begin{center}
\centering
\includegraphics[width=0.68\textwidth]{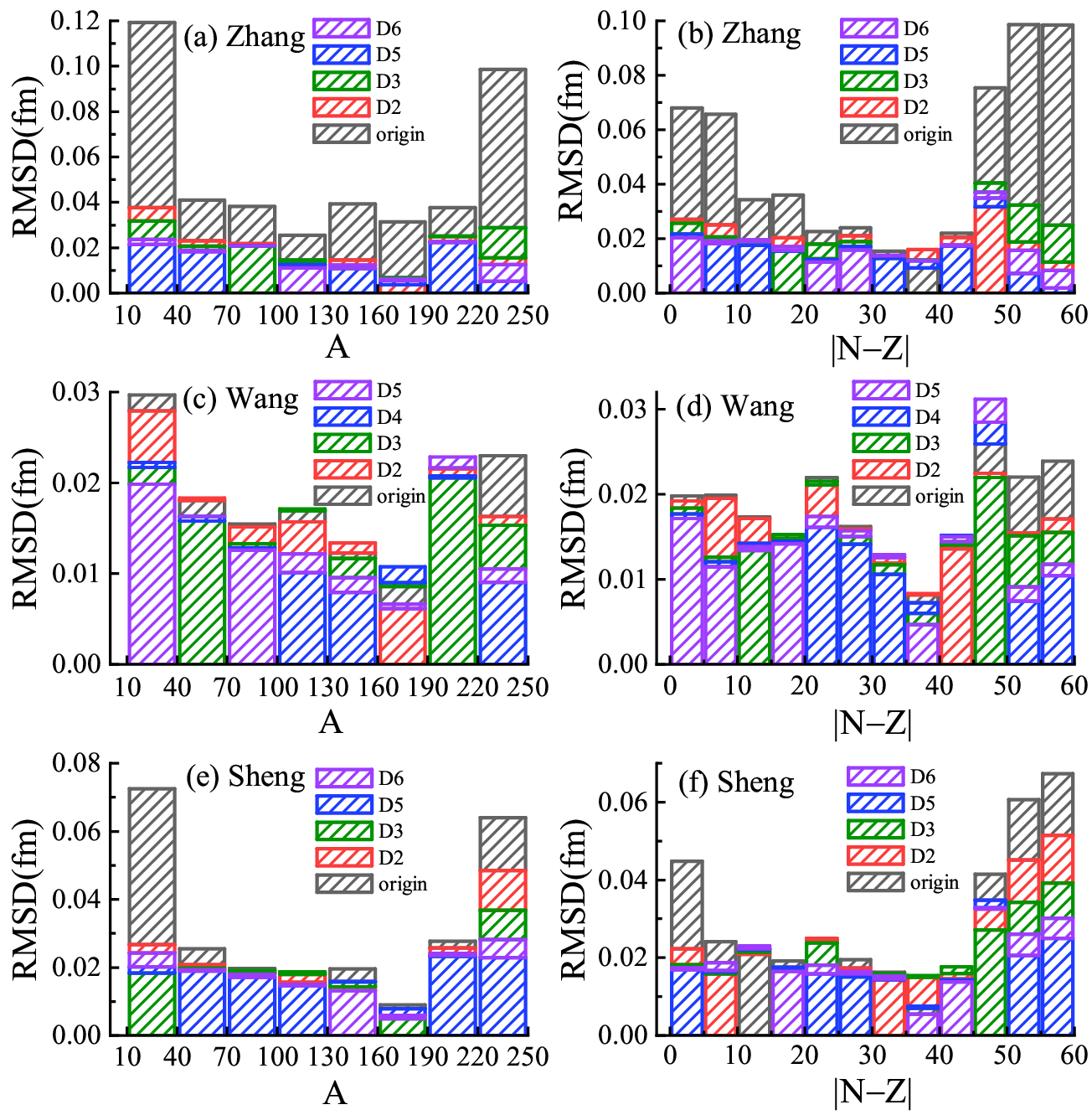}
\vspace{0.5mm}
\figcaption{\label{fig:3}Variation of the RMSDs for nulei with mass A (left column) and isospin asymmetry $|N-Z|$ (right column) in the validation set. In the calculations, three original phenomenological formulas and those corresponding D$i$ models are adopted. From top to bottom, the order is Zhang formula, Wang formula, and Sheng formula.}
\end{center}
\begin{multicols}{2}

By carefully comparing the calculation results of three models with and without abnormal data input, we can see that the dependency behavior of the model calculation with $A$ and $|N-Z|$. This also reflects to some extent that the input of abnormal data does not significantly affect the optimization of these three original formulas by the BNN method. It further indicates that the stronger the physical constraints of the model, the weaker the optimization capability of the BNN method.

\subsection{The influence of hyperparameters on the optimization capabilities of the Bayesian method}

In response to the overfitting phenomenon of the NP94-BNN and ELD13-BNN models in Table 2, we believe that it may be due to the fact that there are only 1013 data in the entire dataset, such that the model itself has less physical information, and the combination of hyperparameters used in the previous calculations may not be the optimal choice. In order to validate the above analysis, we selected five phenomenological formulas to carry out the following simulations: (1) adding hidden layers such that the new FFNN corresponds to the I-J-K-1 and I-J-K-L-1 architectures in Fig.~1, i.e., with two and three hidden layers, respectively; and (2) setting the added hidden layers to have different numbers of neurons. The results are displayed in Table 4. Before simulation, one must guarantee that the width of NNs is strictly larger than the inputs of NN; if the width is smaller or equal to the inputs, then the NN is not a universal approximator~\cite{Lu2017_NIPS}. 

From Table 4, it can be seen that adding hidden layers can significantly improve the predictive capability of BNN for models with overfitting. For example, in the validation set of the NP94-BNN model, the RMSD values calculated by single hidden layer, double hidden layer, and triple hidden layer BNN increased by $32.7\%$, $59.6\%$, and $61.3\%$, respectively, and the calculation accuracy was very close to models such as Sheng-BNN. 
For other phenomenological models with richer physical information, adding hidden layers can improve the optimization ability of BNN, but compared to the optimization capability with only a single hidden layer, there is no significant improvement. For example, for the validation set of the Wang-BNN model, the RMSDs calculated by the three architectures of BNN increased by $16.2\%$, $16.8\%$, and $19.6\%$, respectively. 

In addition, for double hidden layers with different numbers of neurons in each hidden layer, for example, the RMSD values calculated for the BNN of the 6-30-20-1 architecture are similar to those of the BNN with the 6-30-30-1 architecture. The RMSDs calculated for both architectures are only on the order of 0.01~fm, indicating that the number of neurons in each hidden layer has no significant impact on the current research problem. Overall, we believe that the hyperparameter of the number of hidden layers has a significant improvement effect on the optimization of the BNN method.

\end{multicols}
\begin{center}
\tabcaption{ \label{t:4}
RMSDs and relative differences in RMSDs for nuclear charge radii predicted by five BNN models with four difference architectures of NN in the training set and the validation set, respectively.}
\scalebox{0.95}{
\begin{tabular}{c c c c c c c c c c c c c}
\hline \hline
\multicolumn{1}{c}{}&\multicolumn{4}{c}{Training set ($\sigma_{\rm{post}}/\rm{fm}$)} & &\multicolumn{4}{c}{Validation set ($\sigma_{\rm{post}}/\rm{fm}$)} \\
\cline{2-5}\cline{7-10}
\raisebox{1ex}[1ex] {Model} &~$\rm{\uppercase\expandafter{\romannumeral1}}(30)$ & $\rm{\uppercase\expandafter{\romannumeral2}}(30, 30)$ & $\rm{\uppercase\expandafter{\romannumeral2}}(30, 20)$ & $\rm{\uppercase\expandafter{\romannumeral3}}(30, 20, 10)$ & &~$\rm{\uppercase\expandafter{\romannumeral1}}(30)$ &$\rm{\uppercase\expandafter{\romannumeral2}}(30, 30)$ &$\rm{\uppercase\expandafter{\romannumeral2}}(30, 20)$ & $\rm{\uppercase\expandafter{\romannumeral3}}(30, 20, 10)$\\
\hline
NP94-BNN  &0.0315 &0.0218 &0.0229 &0.0218 & &0.0315 &0.0189 &0.0197 &0.0181\\
Zhang-BNN &0.0221 &0.0220 &0.0218 &0.0223 & &0.0188 &0.0191 &0.0187 &0.0199\\
ELD13-BNN &0.0296 &0.0257 &0.0265 &0.0257 & &0.0313 &0.0278 &0.0283 &0.0278\\
Wang-BNN &0.0191 &0.0184 &0.0184 &0.0177 & &0.0150 &0.0149 &0.0141 &0.0144\\
Sheng-BNN &0.0210 &0.0187 &0.0185 &0.0181 & &0.0190 &0.0176 &0.0181 &0.0168\\
\hline \hline
\end{tabular}
}
\end{center}
\begin{multicols}{2}

\end{multicols}
\begin{center}
\scalebox{0.95}{
\begin{tabular}{c c c c c c c c c c c c c}
\hline \hline
\multicolumn{1}{c}{}&\multicolumn{4}{c}{Training set ($\Delta \sigma/\%$)} & &\multicolumn{4}{c}{Validation set ($\Delta \sigma/\%$)} \\
\cline{2-5}\cline{7-10}
\raisebox{1ex}[1ex] {Model} &~$\rm{\uppercase\expandafter{\romannumeral1}}(30)$ & $\rm{\uppercase\expandafter{\romannumeral2}}(30, 30)$ & $\rm{\uppercase\expandafter{\romannumeral2}}(30, 20)$ & $\rm{\uppercase\expandafter{\romannumeral3}}(30, 20, 10)$ & &~$\rm{\uppercase\expandafter{\romannumeral1}}(30)$ &$\rm{\uppercase\expandafter{\romannumeral2}}(30, 30)$ &$\rm{\uppercase\expandafter{\romannumeral2}}(30, 20)$ & $\rm{\uppercase\expandafter{\romannumeral3}}(30, 20, 10)$\\
\hline
NP94-BNN  &43.4 &60.9 &58.9 &60.9 & &32.7 &59.6 &58.5 &61.3\\
Zhang-BNN &55.2 &55.4 &55.8 &54.8 & &60.4 &59.8 &60.6 &58.1\\
ELD13-BNN &30.9 &42.5 &44.0 &44.2 & &16.5 &44.9 &49.3 &48.6\\
Wang-BNN  &14.0 &17.1 &17.1 &20.3 & &16.2 &16.8 &21.2 &19.6\\
Sheng-BNN &34.4 &41.6 &42.2 &43.4 & &34.7 &37.8 &36.0 &40.6\\
\hline \hline
\end{tabular}
}
\end{center}
\begin{multicols}{2}

\end{multicols}
\begin{center}
\centering
\includegraphics[width=0.7\textwidth]{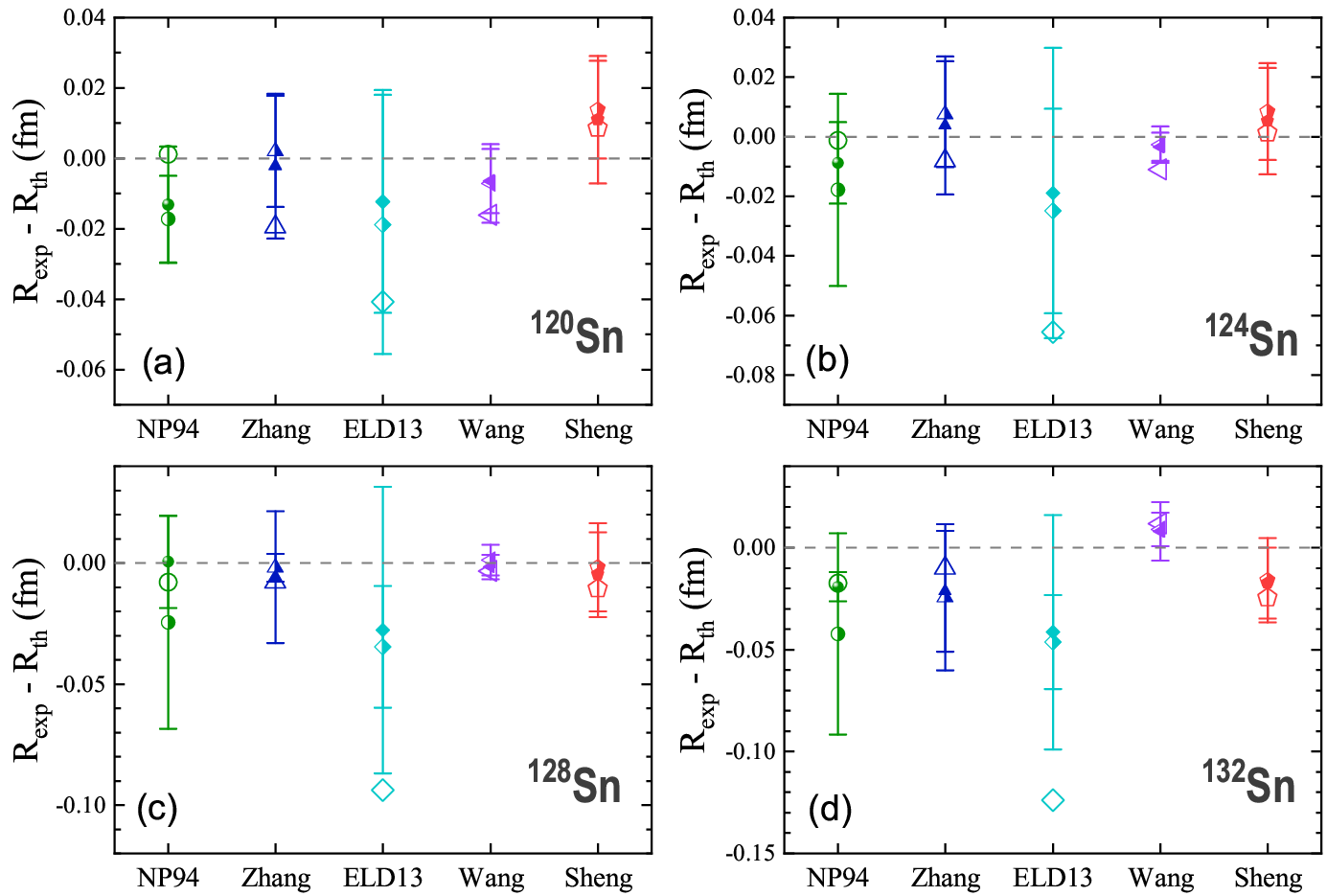}
\vspace{1mm}
\figcaption{\label{fig:4}
 Optimization results of two BNN models for nuclear charge radii of ${}^{120,124,128,132}$Sn isotopes. Comparing to the calculation results of the original phenomenological formulas (hollow symbols), the optimization results of the BNN models for the two architectures 6-30-1 and 6-30-30-1 are represented by half-filled and solid symbols, respectively.}
\end{center}
\begin{multicols}{2}

Figure 4 shows the optimization results of the nuclear charge radii of ${}^{120,124,128,132}$Sn isotopes, using BNN architectures of 6-30-1 and 6-30-30-1, represented by half filled and solid symbols, respectively. As a comparison, we also show the calculation results using 5 original formulas (corresponding to hollow symbols) in the figure. From Fig.~4, it can be seen that for the five models, compared to the optimization of the single hidden layer BNN method, the optimization of the multi-hidden layer BNN has been further improved. Except for the NP94-BNN and ELD13-BNN models, the optimization results of the other three models are close to or less than a level of 0.02~fm. A more careful comparative analysis can lead to the following conclusion: considering the hyperparameter effect of adding hidden layers, the BNN optimization can significantly improve the prediction stability of nuclear charge radius, which provides a reference for other similar works.

\section{Summary\label{Sec:summary}}

In this work, we achieved better predictions for nuclear charge radii with an approach combining the phenomenological formulas with the Bayesian neural network (BNN). We show that as the physical information contained in phenomenological formulas increases, such as Casten factor, shell corrections, isospin asymmetry, pairing, and odd-even staggering effects, the optimization capability of BNN becomes less significant. A suitable combination of the BNN inputs should be carefully selected when predicting and refining the nuclear structure properties or reaction information with the BNN approach.

On the other hand, the stronger the physical constraints of phenomenological formulas of nuclear charge radii, the less optimization the BNN can achieve, and the weaker the extrapolation capability of the BNN, which means the robustness of the Bayesian approach is relatively weakened.

After adding hidden layers for the NN, the optimization capabilities of the BNN approach can significantly improve the prediction stability of nuclear charge radii and successfully overcome the overfitting of the models with less physical information. In addition, changing the number of neurons contained in the hidden layer has a relatively weaker impact on the optimization capability of BNN, which indicates that it is necessary to choose hyperparameters reasonably and carefully while using the Bayesian optimization approach.

\acknowledgments{The authors are grateful to Jian Liu, Xiao-Xu Dong, and Jun-Xu Lu for fruitful discussions and suggestions. This work was supported by National Natural Science Foundation of China (Grants No. 12275115, No. 11447019).}

\end{multicols}

\vspace{-1mm}
\centerline{\rule{80mm}{0.1pt}}
\vspace{2mm}

\begin{multicols}{2}


%

\end{multicols}

\end{document}